\newcommand{\tsecompldate}{2nd November 2007}
\documentclass[11pt]{article}

\usepackage{amsmath,amssymb,amscd}

\usepackage{times} 

\usepackage{graphicx}

\newcommand{\tnote}[1]{} 
\newcommand{\tpre}[1]{} 
\newcommand{\tprenote}[1]{} 

\newcommand{\href}[2]{#2}
\newcommand{\eprint}[1]{\texttt{#1}}

\newcommand{\tref}[1]{(\ref{#1})}

\renewcommand{\tpre}[1]{#1}

\newcommand{\bea}{\begin{eqnarray}}
\newcommand{\eea}{\end{eqnarray}}
\newcommand{\beq}{\begin{equation}}
\newcommand{\eeq}{\end{equation}}
\newcommand{\nnel}{\nonumber \\ {}}

\newcommand{\tsemat}[1]{{\mathbf{\textsf{#1}}}}

\newcommand{\ra}{\rightarrow}

\newcommand{\dprime}{{\prime\prime}}

\newcommand{\Amat}{\tsemat{A}}

\newcommand{\Mmat}{\tsemat{M}}

\newcommand{\Etilde}{\frac{p_r}{p_p}E}
\newcommand{\Ktilde}{\frac{p_r}{p_p}\kav}

\newcommand{\taverage}[1]{\langle #1 \rangle}
\newcommand{\kav}{\taverage{k}}
\newcommand{\ksqav}{\taverage{k^2}}
\newcommand{\efunc}{\omega}

\newcommand{\Sone}[2]{\mathbf{S}_{#1}^{(#2)}}
\newcommand{\Stwo}[2]{\frak{S}_{#1}^{(#2)}}

\newcommand{\rhoexp}{\taverage{\rho}}

\begin{document}

\renewcommand{\thefootnote}{\fnsymbol{footnote}}


\begin{center}
{\Large\textbf{Randomness and Complexity in Networks\footnote{Based on talk given at the workshop on
``Stochastic Networks and Internet Technology'',
Centro di Ricerca Matematica Ennio De Giorgi, Matematica nelle Scienze Naturali e Sociali, Pisa, 17th – 21st September 2007.
 [\texttt{Imperial/TP/07/TSE/4}, \eprint{arXiv:0711.0603}, \tsecompldate]}}}\tnote{tnotes such as this not present in final
version} \\[0.5cm]
 {\large T.S.\ Evans\footnote{\href{http://www.imperial.ac.uk/people/t.evans}{\texttt{http://www.imperial.ac.uk/people/t.evans}}  } }
 \\[0.5cm]
 \href{http://www.imperial.ac.uk/research/theory}{Theoretical Physics},
 Blackett Laboratory, Imperial College London,\\
 South Kensington campus, London, SW7 2AZ,  U.K.
\end{center}

\begin{abstract}
  I start by reviewing some basic properties of random graphs.  I then consider the role of random walks in complex networks and show how they may be used to explain why so many long tailed distributions are found in real data sets.  The key idea is that in many cases the process involves copying of properties of near neighbours in the network and this is a type of short random walk which in turn produce a natural preferential attachment mechanism.  Applying this to networks of fixed size I show that copying and innovation are processes with special mathematical properties which include the ability to solve a simple model \emph{exactly} for \emph{any} parameter values and at \emph{any} time.  I finish by looking at variations of this basic model.
\end{abstract}

\renewcommand{\thefootnote}{\arabic{footnote}}
\setcounter{footnote}{0}

\section{Random Graphs and Random Walks}

\begin{figure}[hbt]
{\centerline{\includegraphics[width=4cm]{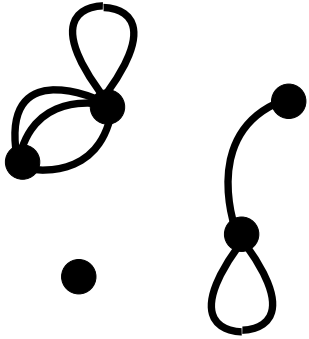}}}
\caption{Example of the type of network considered.  Edges have no directions or values but multiple edges and tadpoles may be  allowed.  Here $N=5$, $E=6$, the largest degree is $k=5$,  the average degree is $\kav = 12/5$,  and the degree distribution is $n(k=0)=1$, $n(k=1)=1$, $n(k=3)=2$, $n(k=5)=1$, with $n(k)=0$ for remaining values of $k$.}
 \label{fexnet}
\end{figure}

In this paper we will focus on undirected graphs where the links between vertices have no values or directions associated with them.  These restrictions can be relaxed in many of the situations examined here but it simplifies the discussion without losing the essential points.  In general we will allow edges to start and end on the same vertex (tadpoles), and for multiple edges between vertices so Fig.\ \ref{fexnet} is a simple example.  Our graphs or networks (the terms are used interchangeably here) consist then of $N$ vertices and $E$ edges.  The number of edges attached to a vertex is the degree, denoted by $k$ and the average degree is $\kav$.  The degree distribution $n(k)$ is the number of vertices with
degree $k$ which when normalised gives $p(k)=n(k)/N$, the probability distribution function.  In section \ref{scopying} we will also use these quantities for a simple bipartite graph. Note that in most cases we imagine creating many copies of a network using some stochastic process.  Thus the averages are often over these ensembles of graphs, not just over all the vertices.  In particular in many cases we will actually be looking at the mean degree distribution and $n(k)$ will be that obtained by averaging over such ensembles.

We will often talk about an object being chosen 'randomly'.  To be ore precise what is meant is that the object is chosen with a uniform probability distribution from the set of similar objects, and it should be obvious from the context what this set is.  For instance if we choose a random vertex of a graph, what is meant is that the probability of choosing a given vertex is simply $1/N$.

\subsection{Random Graphs}\label{ssrndgraphs}

The Classical or Erd\H{o}s-R\'{e}yni Random Graphs may be defined in one of two ways. Either for every pair of distinct vertices add a single edge with probability $p = E/N$, otherwise no edge is added.  Alternatively add $E$ vertices between randomly chosen vertex
pairs.  No difference for large N when sparse $2E/N = \kav \sim O(1)$, similar to the difference between canonical and microcanonical ensembles in statistical mechanics, and then we find a Poisson degree distribution.

Generalised random graphs are graphs which have a given degree distribution $p(k)$ but which are otherwise fixed.  These may be created with the Molloy-Reed construction \cite{MR95,MR98} in which each vertex $a$ is attached to $k_a$ stubs (half an edge), where $k_a$ is drawn from the given distribution $p(k)$.  Then pairs of stubs chosen at random (uniformily) are connected.  Alternatively one may create a graph in any way one likes with the desired $p(k)$ and then use Maslov-Sneppen rewiring \cite{MS02} to randomise graph.  Such generalised random graphs have a given $p(k)$ but otherwise their properties are completely random. In particular the properties of all vertices are the same. For any given source vertex, the properties of neighbouring target vertices will be independent of properties of the source vertex.

This means that random walks on generalised random graphs are particularly simple.
However the existence of an edge does mean that degree distribution of neighbours is not simply $p(k)$
because the higher the degree of a vertex the more likely you are to arrive at that vertex, given there is no correlation between vertices.  Thus the probability that the neighbour of a vertex with degree $k_i$ has degree $k_n$ is given by
\beq
 p(k_n | k_i) = \frac{k_n}{\kav} p(k_n) \, .
 \label{pkneighbour}
\eeq

Let us use this to find the length of random walks on random graphs.  Suppose we follow a random walk where we never go back
along the edge we just arrived on.  Then for infinite graphs $N \ra \infty$ our random walks always end if  $\kav<2$ since we must have arrives on one edge but this leaves 'less than one edge' to continue the walk, i.e.\ sometimes there will be an edge but sometimes not.  This must also mean that we do not have an infinite sized component, no GCC (giant connected component).  On the other hand walks never end and we do have a GCC if $\kav >2$.  The transition to a phase where the GCC exists is at $z=1$  where \cite{NSW01,DMS03a,FFH05,MR95}
\beq
 z(t) := \frac{\kav   }{\kav} -1 = (E-1) F_2(t) \, .
  \label{zdef}
\eeq
In fact all global properties depend on same ratio of second and first moments, $z$ \tref{zdef}, for instance GCC size, the component distribution, and average path lengths.

Let us use the calculation of the average path length in generalised random graph to illustrate these ideas (following Fronczak et al.\ \cite{FFH05}).  Let $p_{ij}$ be the
probability that a random walk in which one never returns along last step taken, starting at vertex $i$, passes through vertex $j$ at
least once after $x$ steps.  The number of different walks of length $x$ from $i$
to $j$ \textit{if no loops} is  $W(i,x) = k_i (k_n-1)^{x-1}$. The probability of \textit{not} arriving at $j$ on any
one step is just $1- k_j/(2E)$.  So the probability that a random walk does not arrive at $j$ after $x$ steps is
\beq
p_{ij} = \left(1- \frac{k_j}{2E} \right)^{W(i,x)}
 \approx \exp\{ - \frac{k_ik_j}{2E} z^{x-1} \} \, .
\eeq
The probability that walker first arrives after $x$
steps is then simply $p_{ij}(x-1) - p_{ij}(x)$ so the average path length from $i$ to $j$ is then
\beq
l_{ij} = \sum_{x=1}^{\infty} x[p_{ij}(x-1) - p_{ij}(x)] =
 \sum_{x=0}^{\infty} p_{ij}(x) \, .
\eeq
This gives the average path length between any two randomly chosen vertices as
\beq
 \langle l \rangle = \frac{-2 \langle \ln(k) \rangle + \ln(E)  -
 \gamma_E}{\ln(z)} + \frac{3}{2}
 \, , \qquad \gamma_E \approx 0.5772 \;\; .
 \label{LCdist}
\eeq

Calculations like this above work because
of the lack of correlations between vertices in such random graphs and because for for large sparse graphs the graphs are basically trees with no loops.
These can be reasonable approximations for many models
and perhaps for a few real graphs too. Otherwise we may use generalised random graphs as a \textit{null model} against which we can compare other networks. Often these calculations are exact for closely related \emph{Urn
models} (see later discussions).

\subsection{Random Walks}\label{ssrndwalks}

Given the inspiration from the analysis of generalised random graphs and Eqn.\ \ref{pkneighbour} in particular, let us now consider random walks as a tool for networks of all types. Random Walks are the extreme alternative to the use
of Shortest Paths.  One would normally use random walks to discuss mean first passage time etc and these are related to eigenvalues/vectors a transfer matrix defining a Markovian diffusion process on the graph.  They are used for calculations of generalised random graphs (as seen above), sampling graphs \cite{HHMN00,GMS04,OS04,Yang05,SG07,CT07,SG07}, community detection \cite{PL05a,PL06} and the natural creation of scale-free networks \cite{Vaz03,SK04,ES05,SLJ06,SLJO07a,SLJO07b,CC07}.

Consider an unbiased random walk, where one treats all edges as equal (including the one used to arrive at the current vertex).  Used to sample networks, so as a tool to search the vertices of a network, then vertices are visited with probability roughly $p_\mathrm{visit}(k) \approx k p(k) / (2E)$.  This means that \textit{hubs} (large degree nodes) are found very quickly so the tail of the degree distribution may be easily estimated.  Other biased walks are also possible but these do not share the
same special properties e.g.\ can sample vertices equally if slowly \cite{OS04,Yang05,SG07}.

When we think of random walks as a diffusion process then we define a transition matrix $\Mmat$ in terms of the adjacency matrix, $\Amat$.  If we define $A_{ij}$ to be the edge value \emph{from} vertex $j$ \emph{to} vertex $i$, then for an unbiased random walk we want the probability of moving \emph{from} vertex $j$ \emph{to} vertex $i$ to be the entry $M_{ij}$ where
\beq
M_{ij} := \frac{A_{ij}}{k_j} \, .
\eeq
If we suppose the number of random walkers at vertex $i$ at time $t$ is $v_i(t)$ then we have to solve the matrix equation $v(t) = \Mmat(t) v(0)$, a simple Markov process.  For an undirected graph ($A_{ij}=A_{ji}$) the solution is simple
\beq
 v_i(t) = c_1 \frac{k_i}{2E} + \sum_{n=2}^{N} c_n (\lambda_n)^tu_i^{(n)}
  \label{simplediff}
\eeq
where the $n$-th eigenvector of the Markovian  matrix $\Mmat$ is $u_i^{(n)}$ associated with eigenvalue $\lambda_n$.  Since the adjacency matrix has non-negative entries, if we assume the graph is connected then from the Perron-Frobenius theorem we know the eigenvectors are real and ordered as $1 = \lambda_1 > |\lambda_2| \geq \ldots \geq |\lambda_j| \geq \ldots \geq 0$.  The largest eigenvalue is equal to one, indicating a single stable long time solution specified by the first eigenvector which is $u_i^{(n=1)} =k_i/(2E)$ for our undirected case.
Note that adjacency matrix eigenvectors and eigenvalues have
no obvious physical interpretation, nor do they have any simple relationship to those of the Markovian $\Mmat$ matrix.

Given the existence of such eigenvectors, it is natural to try some sort of spectral analysis.
The solution \tref{simplediff} shows us that the eigenvectors can sometimes be interpreted as informing us about poorly connected regions \cite{ESMS03}.  However they are generally a poor way of determining community structure.  More useful are approximation schemes where one uses just a few eigenvectors associated with the largest eigenvalues to reduce the dimension of the matrix, $\Mmat$, from $O(N^2)$ to some approximate $O(N)$ structure.  This is the basis of Principal Component Analysis and Singular Value Decomposition (e.g. see \cite{LMV00,WRR03}).

One may also use these eigenvectors to provide a ranking.  In this case one defines the ranking value of vertex $i$ to be the $i$-th entry of the first eigenvector of some Markovian matrix. For our unbiased walk $\Mmat$ we found this was imply the degree of an undirected graph.  However one can easily considers more complicated random walks.  For instance
\beq
 M_{ij} := (1-p_v) \frac{A_{ij}}{k_j} + p_v \frac{1}{N}
\eeq
is equivalent to a walk where one follows a randomly chosen edge out of a vertex with probability $(1-p_v)$ but otherwise one jumps to a random vertex, essentially starting a new random walk.  Such variations are the basis for PageRank used by Google \cite{BP98b}.

So why is a random walk so useful? In all these examples we are exploiting the fact that a random walk probes global structure of network but uses
only local information.  This is computationally efficient for computer algorithms which are searching the whole graph.  However it also exactly the same feature that is required in the real world by those creating or using network. A process involving only local information is much more
likely to occur naturally i.e. no external influence needed.  The author of a web page knows only a small neighbourhood of existing web page.  So when adding a link to their web page an author will have surfed a local neighbourhood in a way that, when we average over the behaviour of many such authors,  might be statistically indistinguishable from a random walk.  This suggests that we might be able to use random walks to answer a much deeper question: Why do so many networks have long tailed degree distributions?

\subsection{Long Tails of Growing Networks}\label{sslongtails}

\begin{figure}[hbt!]
\begin{center}
\includegraphics[width=7cm]{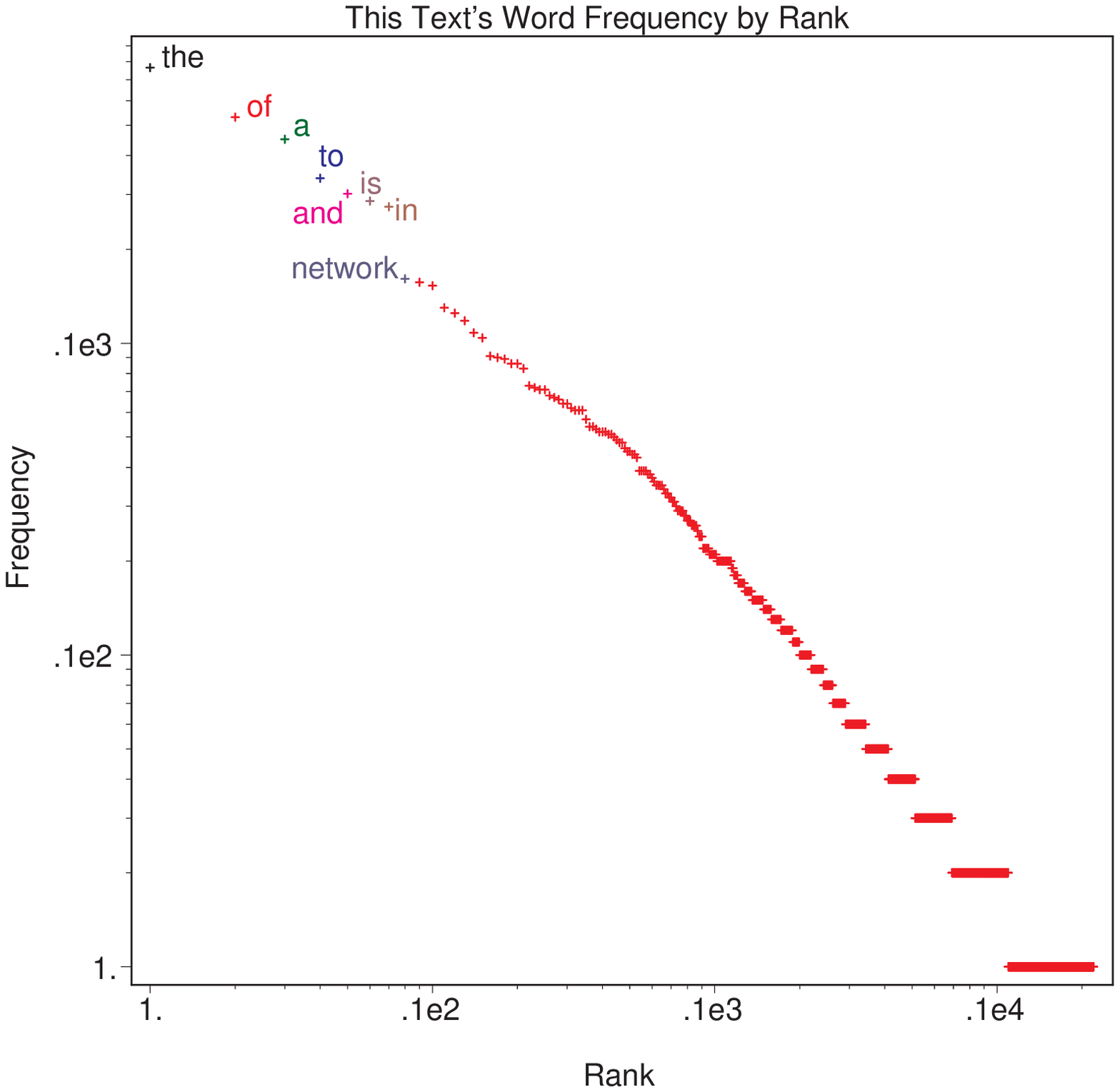}
\includegraphics[width=7cm]{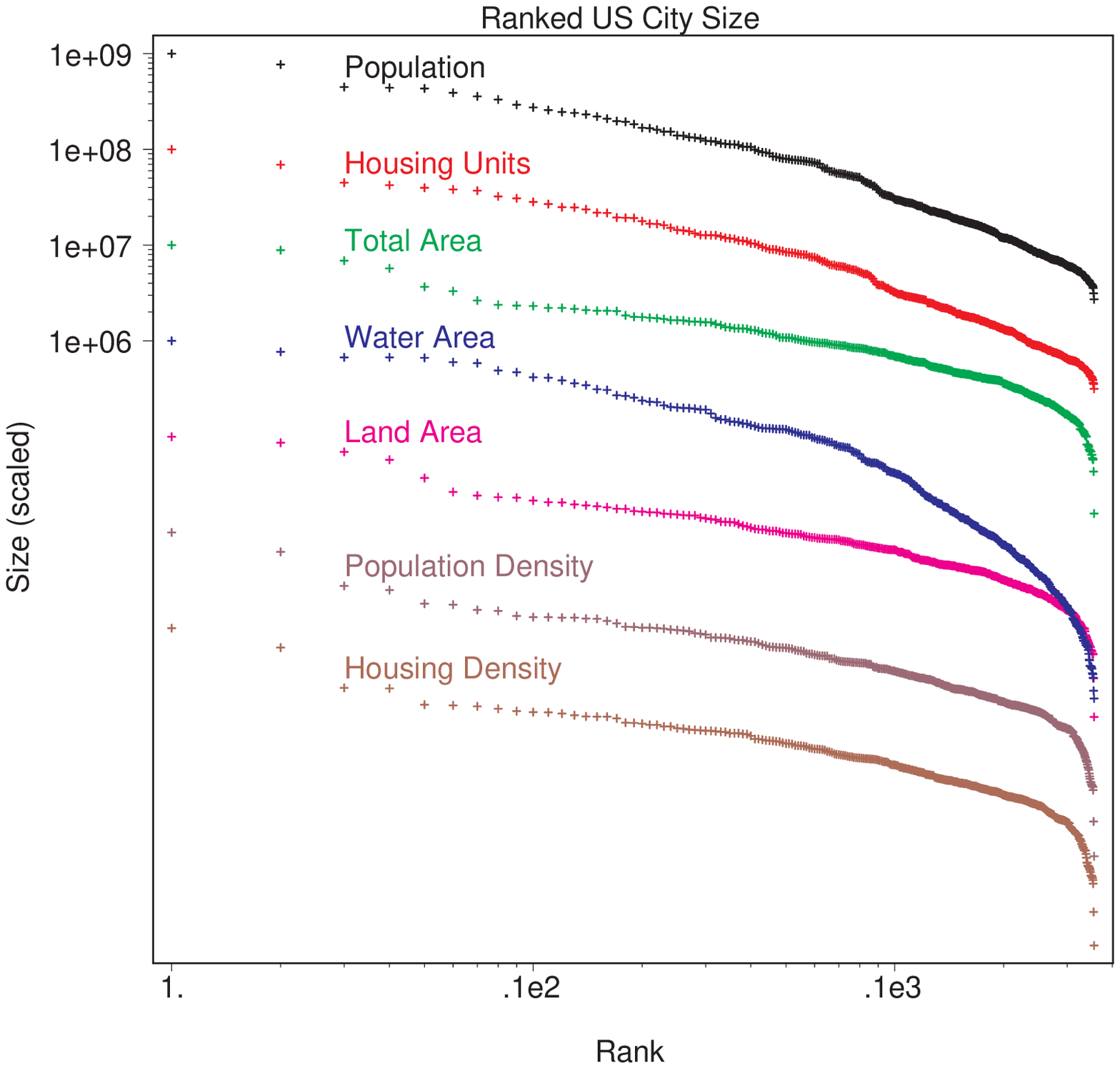}
\end{center}
\caption{Two classic examples of long tailed distributions in real data sets.
On the left is a log-log plot of
word frequency against rank for the review \cite{Evans04}  ---  ``the'' is the most
common word, ``network'' is eighth. On the right is a log-log plot
of the size of US ``Metropolitan Areas'' (measured in various ways) against rank, all data scaled relative to
the largest and by a power of ten relative to the next curve (for
visualisation purposes). Most show some
evidence of a simple power law.}
 \label{flongtaildata}
\end{figure}

\begin{figure}[hbt!]
\begin{center}
\includegraphics[width=8cm]{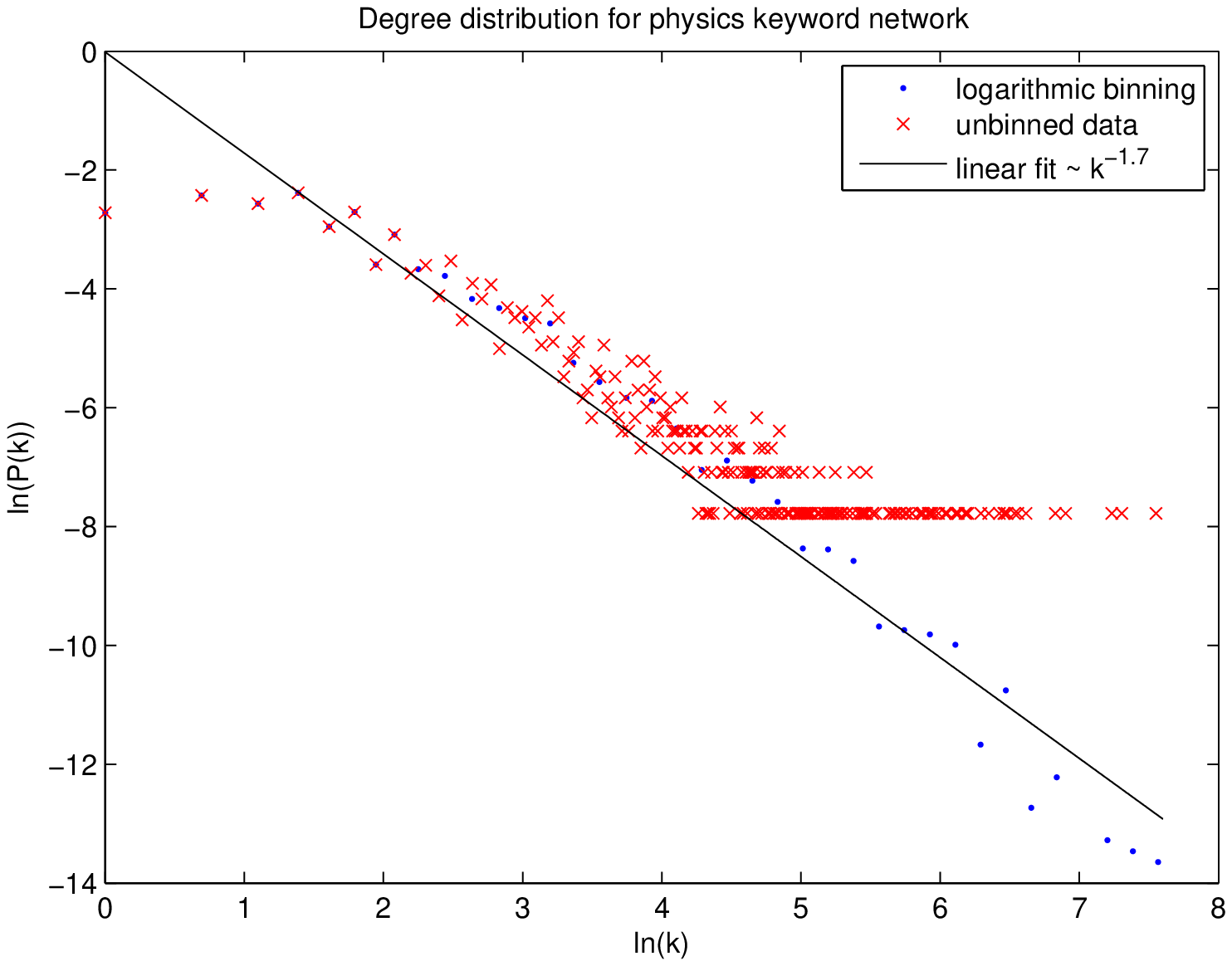}
\includegraphics[width=6cm]{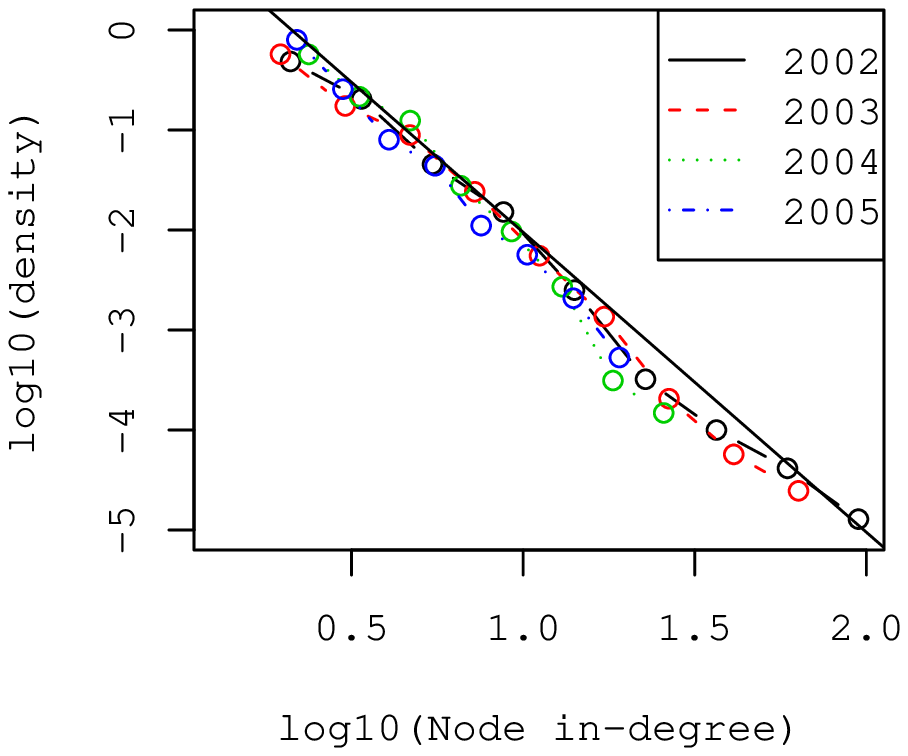}
\end{center}
\caption{Two examples of networks possessing long tailed distributions.  On the left is a plot (with and without logarithmic binning) of the degree distribution taken from a network derived from the publications of the Physics department of Imperial College where the vertices are key words \cite{EHW06}.  On the right is the distribution of the in-degree of the logged in user
graph for data taken from the website
\texttt{gallery.future-i.com} where users may publish pictures (taken from \cite{AEH07}).}
 \label{flongtailnets}
\end{figure}

Long tailed distributions are common in data sets, Fig. \ref{flongtaildata} shows two examples.
This is equally true for networks, for example Fig. \ref{flongtailnets}, where long tailed distributions indicate the presence of large \emph{hubs}, vertices of high degree.
Lattices,  Small World networks \cite{WS98}, and classical random graphs have no hubs.  Only a long tailed degree distribution has hubs such as a power law e.g.\ if $n(k) \sim 1/k^3$ then with $N=10^6$ $\kav =4$ the largest vertex will be of degree around 2500 while for a classical random graph it will be about 17.

The standard model used to produce such long tailed distributions has a long history being discussed by Yule \cite{Yule24}, Simon \cite{Simon55,Simon57,BE01a} and Price \cite{Price65,Price76} amongst others.  However it was put into the context of networks by Barab\'{a}si and Albert \cite{BA99} who suggested the following algorithm.  Suppose at time $t$ we add a new vertex to an existing graph. We then attach one end of each of $\kav$ new edges to the new vertex and attach the other end to an existing vertex in the network.  These existing vertices are chosen with \emph{preferential attachment}, that is they are selected with probability $k/(2E)$ so that the ``Rich get Richer'', vertices with many edges are favoured to gain more.  The result is that after a relatively short time a long tail appears with an asymptotic form $n(k) \sim k^{-3}$. Subsequent work shows that one can produce such power law tails in many ways.  By choosing to mix in some random attachment or by not adding a vertex at every time step, one may produce any power from two to infinity \cite{KRL00,KRR01,BAJ99,DMS00,CNSW00,KR01,KK01}.  Growth is not essential as one might use an appropriate Hamiltonian and some rewiring scheme \cite{BM03}.  Even the network picture is not needed as the older works of Yule, Simon and Price show.  However, what is clear is that the attachment probability \textit{must}  be exactly linear in degree (for large degrees) otherwise non-power law degree distributions appear  \cite{KRL00}.  This begs the question, why do so many real systems appear to be attaching new edges with perfect linear attachment probabilities?  It is easy for a computer to generate the attachment probabilities of $k/(2E)$ but in reality in most networks each node has only relatively limited and local information.  That is the $E$ in the normalisation of the Barab\'{a}si and Albert algorithm is unknown!

The most natural solution to the frequent appearance of long tails in networks is to imitate the behaviour and knowledge available at a vertex in most problems, that is use only local information \cite{Vaz03,SK04,ES05,SLJ06,SLJO07a,SLJO07b,CC07}  So again imagine that we are adding a new vertex to an existing network and we attach this new vertex to   $\kav$ new edges.  The other ends of these new edges we attach to existing vertices which are found by executing a random walk on the existing network.  As we discussed in section \ref{ssrndgraphs}, under fairly general circumstances the walk will arrive at a new vertex with probability proportional to the number of ways of arriving at that vertex, namely its degree, as expressed in \tref{pkneighbour}.  Thus with the simplest algorithm, i.e.\ with only local knowledge used, we generate attachment probabilities proportional to the degree.  Hence we find a long tailed power-law distribution.

\begin{figure}[!tb]
\begin{center}
 \includegraphics[width=7cm]{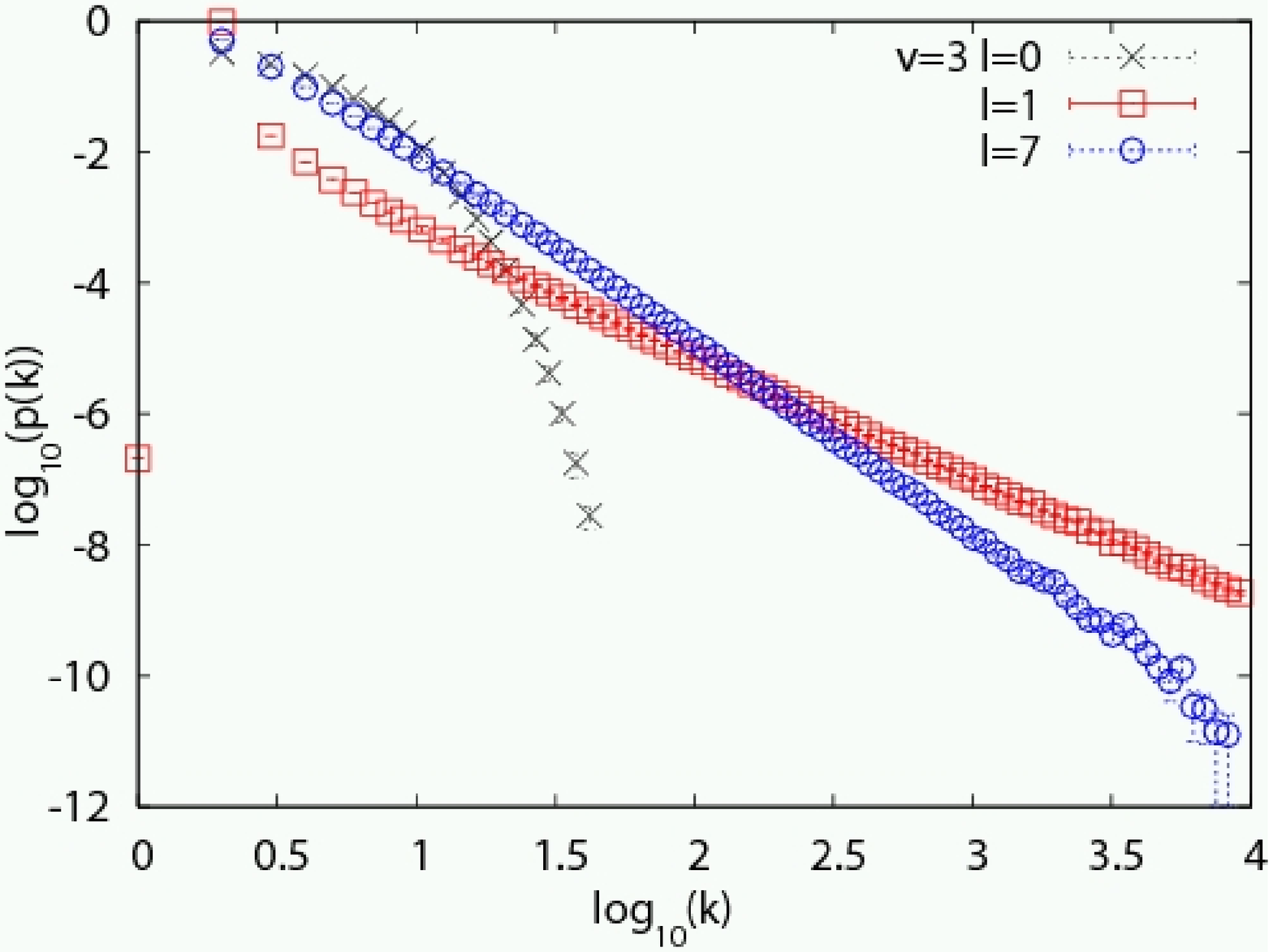}
 \includegraphics[width=7cm]{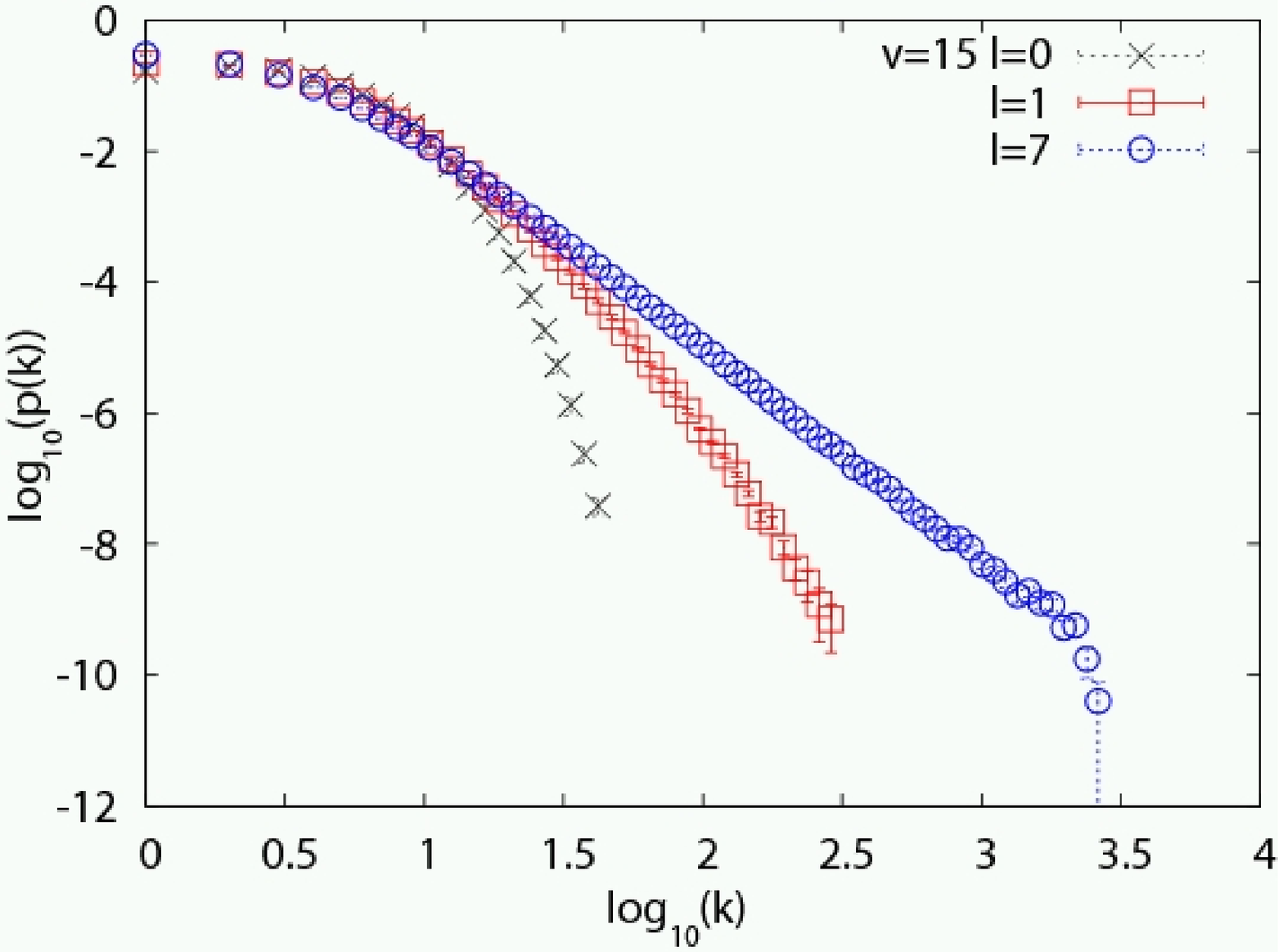}
 \end{center}
\caption{Log-log plots of degree distributions $p(k)$ for $N=10^6$
degree networks generated by random walks started from a randomly
chosen vertex, with one vertex and two edges added at each time step. In
each graph, the results are shown for average walk lengths of zero
(crosses), one (squares) and seven (circles) steps, with data averaged
over 100 runs. In the left diagram, the walk length is fixed and for
each added a new walk is started from a randomly chosen vertex. The algorithm used for the bottom figure has variable numbers of edges and variable walk length. Multiple edges
are allowed but are rarely created. Taken from \cite{ES05}.}
 \label{fESresults}
\end{figure}

One might be concerned that one needs to make a long walk, say of order the average shortest distance between vertices or of order the graph diameter, before we get effective preferential attachment.  In fact as Fig.\ref{fESresults} and the more extensive results of \cite{ES05} show, as long as some of the walks are one step long then a power-law distribution is appears. Walks of length two or more in length produce very similar power laws.  What this suggests is that it is that the important length scale is the degree correlation length and that this is often going to be less than one for many examples.

\begin{figure}[!htb]
\begin{center}
 \centering\includegraphics[width=7cm]{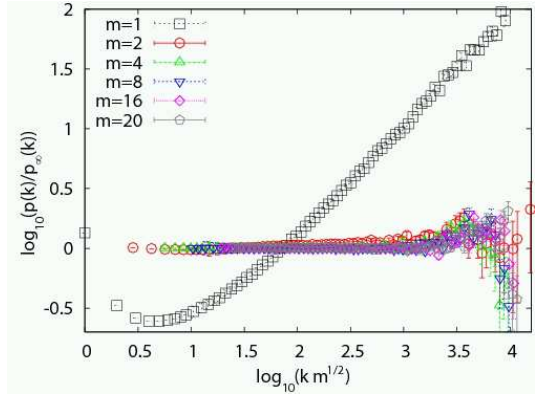}
 \end{center}
\caption{The degree distributions normalised against the appropriate
large network solution $p_\infty(k)$ for fixed number of edges
$E=2\times 10^6$,  with one vertex ($\epsilon=1$) added at each time
step ($\epsilon=1$) but with the average degree ($\kav = 2m$) varied.
Plotted against $\log_{\mathrm{10}}(km^{1/2})$ to take account of
large scale finite size effects since here the number of vertices is
$N \propto 1/m$. For random walks starting from a random vertex for
every new edge, of fixed length $l=7$ and averaged over 100 runs.
Note that the tree graphs formed when $m=1$ (squares) are the only
ones showing a strong deviation from the expected cubic power law,
but they still show good power law behaviour with a power of $\gamma
\approx 2.0$.  Taken from \cite{ES05}.}
 \label{fESmvaryres}
\end{figure}

In a similar way Fig. \ref{fESmvaryres} illustrates that the average degree of the graph makes little difference to the power law tail provided it is bigger than two.  The case of $\kav=2$ produces a tree graph so is a special case but it actually has an even longer tail.

So the random walk algorithm \cite{SK04,ES05} is extremely robust, producing power laws almost whatever one does. Different starting points for the walks, varying the length of the walks changing the number of edges added per vertex still gives power-law tails.  The value of the power is not, however, universal so in that sense it does \textit{not} behave like a critical exponent.  Powers can easily vary by 10\% or 20\% from that expected from an equivalent Barab\'{a}si-Albert algorithm. Nevertheless  this is a self-organised method in that the algorithm uses the structure of the graph to generate its own growth and this still drives it to a power law degree distribution.

\section{Copying}\label{scopying}

There is another way of looking at the random walk process used to create a growing graph.  That is the final step of a random walker links the penultimate vertex on its walk to one of its neighbours.  When we add an edge from our new vertex to the final vertex in the walk, we might imagine that what we are actually doing is \textit{copying} the choice made by the penultimate vertex.  Thus if these edges were links between web pages (the vertices), we are saying that both the new web page and the existing web page think the target of their common edges is a web page worthy of note.  Indeed this is the basis of the utility of Google's PageRank method which as we have seen is essentially based on a random walk of the web.  We will now see how this concept of copying can be extended to a wider class of problems.

\subsection{A Simple Model of Cultural Transmission}

Let us focus on the idea of copying and look at a simple model of cultural transmission --- the \emph{Copying Model} \cite{Evans07,EP07eccs06,EP07,EP07eccs07,EPY07}.  Suppose we have a fixed population of $E$ individuals, each of whom can choose one $N$ `artifacts'.  These artifacts have no intrinsic benefit --- they may be the breed of pedigree dog they own, the shoe style they wear, the name of a baby, style of pottery used by that person.  At each time step, one person chosen with probability $\Pi_R$ updates their
choice, that is they choose a new artifact with probability $\Pi_A$.  In fact we will focus on individuals chosen at random while their new artifact choice will be picked in one of two ways.  They can \textit{copy} the choice made by an individual (chosen at random).  Alternatively they pick an artifact at random.  In the second case, if there are very large numbers of artifacts, $N \ra \infty$, then this will be the first time this type of artifact has been chosen so we can think of this process as \textit{innovation}.  Only after both of these choices are made is the actual network updated.  Thus
\beq
 \Pi_R = \frac{k}{E} \, , \;
 \Pi_A = p_r\frac{1}{N} + p_p\frac{k}{E} \, ,
   \;\
   p_p+p_r=1 \, ,
    \; (E \geq k \geq 0) \; .
 \label{PiRPiAsimple}
\eeq

\begin{figure}[hbt!]
{\centerline{\includegraphics[width=7cm]{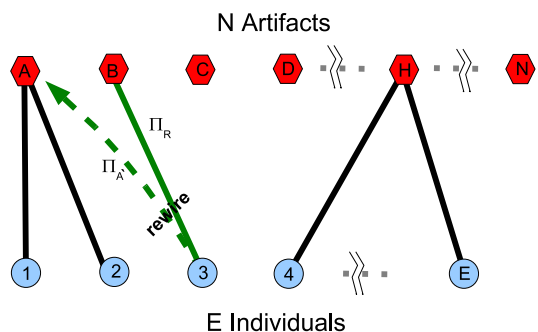}}}
\caption{The Copying Model as a bipartite graph has $E$ `individual' vertices, each
with one edge. The other end of the edge is connected to one of $N$
`artifact' vertices. If the degree of an artifact vertex is $k$ then
this artifact has been `chosen' by $k$ distinct individuals. At each
time step a single rewiring of the artifact end of one edge occurs.
An individual is chosen (number 3 here) with probability $\Pi_R$
which gives us the source artifact (here B).  At the same time
the target artifact is chosen with probability $\Pi_A$ (here
labelled A). After both choices have been made the rewiring is
performed (here individual 3 switches its edge from artifact B to
A).}
 \label{fCopyModel3}
\end{figure}

We can represent this model rewiring of a bipartite network as shown in Fig.\ \ref{fCopyModel3}. In fact the study of networks of constant size has received relatively little attention despite the fact that many systems will eventually reach a constant size or at least change size only slowly.

\begin{figure}[htb!]
{\centerline{\includegraphics[width=7cm]{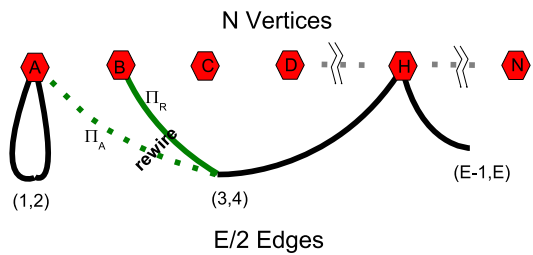}}}
\caption{A Molloy and Reed \cite{MR95} projection of the copying model bipartite graph onto an undirected unipartite graph.  Here individual vertices numbered $(2i)$ and $(2i-1)$ are paired to give the unipartite graph edge labelled $(2i-1,2i)$. The rewiring event shown is the equivalent to that shown in  Fig. \ref{fCopyModel3} for the bipartite graph. }
 \label{fCopyModel2und}
\end{figure}

This bipartite graph may seem to be a trivial network but a projection onto a graph of just the artifact vertices, see Fig. \ref{fCopyModel2und}, is just an implementation of the Molloy-Reed projection \cite{MR95}.  Thus the Copying Model captures the degree distribution of a fixed size unipartite graph undergoing rewiring, which has been studied elsewhere in several ways \cite{WS98,BCK01,DMS03,DM03,PLY05,XZW05,OTH05,OYT05,OYT06}

This Copying Model may seem naive but it has been used on several different data sets:
transmission of cultural artifacts such as pottery designs, dog
breed and baby name popularity
\cite{Neiman95,BS03,HB03,HBH04,BHS04,BS05}.  The copying and innovation are familiar in other contexts, as inheritance and mutation in diversity of
genes \cite{KC64,CK70} or species \cite{CCHJ02,AJ05,LJ06,LJ06a,LJ06b,FC06d,LJ07,Jen07}, or as inheritance and the effect of New Immigrants on the distribution of family names in constant populations \cite{ZM01}.
One may also relate this to models of
language evolution \cite{SCES07} and to variants of the Minority Game \cite{ATBK04}.

\begin{figure}[htb!]
{\centerline{\includegraphics[width=8cm]{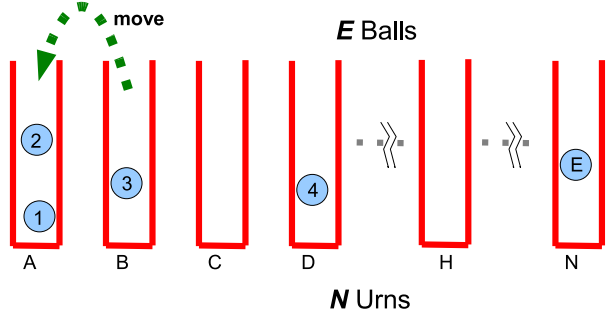}}}
\caption{The Urn model representation of the Copying Model.  The rewiring event shown is the equivalent to that shown in Fig. \ref{fCopyModel3} for the bipartite graph.}
 \label{fCopyModel3urn}
\end{figure}

There is also a close relationship between this copying model and other models of statistical physics models.  This can be translated into the language of Urn models \cite{GBM95,GL02,OYT05,OYT06} as shown in Fig. \ref{fCopyModel3urn}, and is related to some
variations of the Backgammon or Balls-In-Box models used for glasses
\cite{Ritort95,BBJ97}, simplicial gravity \cite{BBJ99} and wealth
distributions \cite{BJJKNPZ02}. The closest zero range process \cite{EvansMR00,EH05,PM05} to the copying model discussed here is the variant with a `misanthrope' process on a fully connected geometry.  Voter Models \cite{Liggett85,SR05}, when played on complete graphs, are just the $N=2$ limit of the case considered here.

\subsection{Mean Degree Distribution}

The mean field approximation is very accurate for many models because of the low
vertex correlations.  However can the mean field equation ever be exact?
The answer is yes but only for special attachment and removal probabilities.

The evolution of the degree distribution is given by
\begin{eqnarray}
\lefteqn{n(k,t+1) - n(k,t)}
 \nnel
 &=&   n(k+1,t) \Pi_R(k+1,t) \left( 1- \Pi_A(k+1,t) \right)
 \nnel
 &&
     - n(k,t)   \Pi_R(k,t)   \left( 1- \Pi_A(k,t)   \right)
 \nnel
 &&
     - n(k,t)   \Pi_A(k,t)   \left( 1- \Pi_R(k,t)   \right)
 \nnel
 &&
     + n(k-1,t) \Pi_A(k-1,t) \left( 1- \Pi_R(k-1,t) \right)
   \, ,
     \nnel
 &&
 \qquad \qquad \qquad \qquad \qquad \qquad (E \geq k \geq 0) \, .
   \label{neqngen}
\end{eqnarray}
Note that the factors of $(1-\Pi)$ are invariably ignored in the literature yet they are essential if we are to enforce the boundary condition that $n(k)=0$ if $k>E$.  These  $(1-\Pi)$ terms take account of processes where the edge is removed and then reattached to the same artifact.

It is implicit that we are taking an ensemble average over many runs of our system.  Thus problems arise when we can deal with the normalisations of our probabilities.  For
instance if we have attachment or removal probabilities
of the form $(k^\beta/z_\beta)$ then the normalisation $z_\beta$ depends on the particular configuration $n(k)$ of each contribution to the ensemble.  That is in general we can factorise as needed:
\beq
 \langle n(k,t) \frac{k^\beta}{z_\beta(t)} \rangle
 \neq
  \frac{\langle k^\beta n(k,t) \rangle }{\langle z_\beta(t) \rangle}
 \; .
 \label{nonlinprob}
\eeq
The only two cases where the mean field approximation is
exact, where we have equality in
\tref{nonlinprob}, is when $\beta=0$ or $\beta=1$.  This is because then the normalisations are invariants of the system, $N$ and
$E$ respectively. The most general choice for $\Pi_R$ and $\Pi_A$
satisfying these criteria is the simple copying and innovation probabilities of \tref{PiRPiAsimple} and we will now restrict ourselves to this case.

It turns out that one can then solve for the mean degree distribution \emph{exactly} for \emph{any} parameter value and \emph{any} time.  The best way is through the generating function $G(z,t)$
\beq
 G(z,t) := \sum_{k=0}^{E} z^k n(k,t) \, ,
 \label{Gktdef}
\eeq
which is like taking a discrete Mellin transform.  The degree distribution and moments are then simple derivatives of the generating function
\bea
 n(k,t) &=& \left. \frac{d^kG(z,t)}{d^k z}\right|_{z=0}\, ,
 \\
 \taverage{k^n} &=& \left. \frac{d^nG(e^x,t)}{d^nx}\right|_{x=0} =
 \left. \left(z\frac{d}{dz}\right)^nG(z,t)\right|_{z=1}  \, .
\eea
This turns our equation \tref{neqngen} into a differential equation
for the generating function,
\bea
 \lefteqn{\frac{b(1+a-c) }{(1-z)}    \left[  G(z,t+1)-G(z,t)\right]
 }
 \nnel
 &=&
  z(1-z)G^{\dprime}(z,t)
 +[c-(a+b+1)z]G^{\prime}(z,t)
 -ab G(z,t) \;,
 \nnel
 &&
 \label{gendifft}
\eea
where the $G^{\prime}$ and $G^{\dprime}$ are single and double derivatives with respect to $z$. The constants $a$, $b$ and $c$ are given by,
\beq
 a = \frac{p_r}{p_p}\kav \; , \qquad
 b = -E \; , \qquad
 c = 1+\frac{p_r}{p_p}\kav-\frac{E}{p_p}.
\eeq
We can exploit the linearity by splitting the generating function into $(E+1)$ eigenfunctions $G^{(m)}(z)$ and eigenvalues $\lambda_m$ ($m=0,1,\ldots,E$):
\begin{equation}
 G(z,t) = \sum_{m=0}^{E} c_m (\lambda_m)^t G^{(m)}(z) \; ,
 \qquad
 G^{(m)}(z) := \sum_{k=0}^{E} z^k \efunc^{(m)}(k) \, .
 \label{Gkmdef}
\end{equation}
The initial conditions
$n(k,t=0)$ fix the coefficients $c_m$.  The eigenfunctions satisfy
\bea
 z(1-z) G^{(m)\dprime} (z)
 + [c-(a+b+1)z] G^{(m) \prime}(z) &&
 \nnel
 - \left[ ab-\frac{(\lambda_m -1)}{1-z} b(c-a-1) \right] G^{(m)}(z)
 &=& 0 \; .
 \label{eq:gendiffm}
\eea
Recognising that this equation is similar to the hypergeometric ODE
we obtain our solution in terms of
the Hypergeometric functions $F = {}_2F_1$ where
\bea
 G^{(m)}(z) &=&  (1-z)^m F(a+m,b+m;c;z)
  \label{Gmresult}
  \\
  \nonumber
 &=& (1-z)^m
 \sum_{l=0}^{E-m}\frac{\Gamma(a+m+l)\Gamma(b+m+l)\Gamma(c)}{\Gamma(a+m)\Gamma(b+m) \Gamma(c+l) (l!)} z^l
\eea
with corresponding eigenvalues,
\beq
\lambda_{m} =
1 -m(m-1) \frac{p_p}{E^2} - m \frac{p_r}{E},
 \qquad 0 \leq m \leq E \; .
 \label{eq:evalues}
\eeq
The eigenvalues satisfy $\lambda_{m} > \lambda_{m+1}$ except for
$p_r=0$ when $\lambda_0=\lambda_1=1$.

\subsection{Exact Equilibrium Solution}

The properties of the hypergeometric function give us the mean degree distribution as a simple ratio of $\Gamma$ functions:
\bea
n(k) &=& A \;
 \frac{\Gamma\left( k + \Ktilde \right) }{\Gamma \left(k+1\right) }
  \frac{\Gamma\left( \frac{E}{p_p} - \Ktilde  - k\right) }{\Gamma \left(E  +1  -  k  \right) }
  \, ,
      \label{pkcomplete}
      \\
      A &:=&
      N\frac{\Gamma\left(\Etilde\right)\Gamma\left(E+1\right)}{\Gamma\left(\frac{p_r}{p_p}(E-\kav)\right)\Gamma\left(\Ktilde\right)\Gamma\left(\frac{E}{p_p}\right)} \, .
\eea
This is similar to the long time solution for growing networks \cite{KRL00,KRR01,BAJ99,DMS00,CNSW00,KR01,KK01} but the second fraction in \tref{pkcomplete} is only found for network rewiring with the correct master
equation, i.e.\ when the factors of $(1-\Pi)$ are included in \tref{neqngen}.  Only approximate solutions were known previous to \cite{Evans07}.
\begin{figure}[htb!]
\centering\includegraphics[width=8cm]{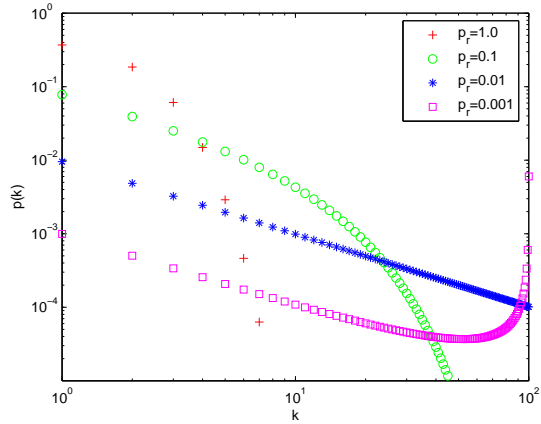}
\caption{The degree probability distribution function
$p(k)=n(k)/N$ for $N=E=100$ and various $p_r$.  From top to bottom at $k=2$ we have: $p_r=1$ (red crosses),
$p_r=10/E$ (green circles), $p_r=1/E$ (blue stars) and $p_r=0.1/E$ (magenta
squares). For $p_r=1/E$ the distribution is almost a pure inverse power law for all
values of $k$.  With $p_r=1$ we have a binomial distribution. Taken from \cite{EP07eccs07}.}
 \label{feqDD}
\end{figure}

For large degree the equilibrium behaviour splits into three regimes.
With a reasonable amount of innovation, $E^{-1} \lesssim p_r \lesssim  (1+\kav)^{-1}$,
the degree distribution is a power law with an exponential cutoff
\beq
 n(k) \approx k^{-\gamma} e^{-\zeta k} \, , \qquad
 \gamma = 1- \frac{p_r\kav}{(1-p_r)} \leq 1 \, , \;
 \zeta = -\ln \left( 1-p_r \right) \, .
 \label{neqdist}
\eeq
The slope $\gamma$ will be indistinguishable from one in data sets as if $\gamma \ll 1$ then the exponential cutoff scale $\zeta^{-1}$ is too small.  This type of solutions is characteristic of a simple copying process in a network of fixed size so it is not surprising to see it appearing in other apparently more complex systems which have copying as part of their fundamental dynamics.  For instance, power laws of one appear in network models of species \cite{LJ06b,FC06d} when the networks are of constant size, at least on average in the long time limit. In these models the copying and innovation processes are inheritance and mutation. Another example, this time from sociophysics \cite{ATBK04}, will be discussed below.

The second region is where randomness dominates, $1 \geq p_r p_r \lesssim  (1+\kav)^{-1}$ so $\zeta^{-1} \lesssim O(1)$.  The degree distribution starts to look more like the binomial distribution which is the limit at $p_r=1$.

The last region appears when in one generation, the time taken to rewire most of the edges once, the edges are likely to be assigned using only the copying process.  This occurs when $E^{-1} \gg p_r \geq 0$.  Now the distribution turns up at $k=E$ and we find almost all individuals are attached to a single artifact --- we have a \emph{condensate} or \emph{fixation}.  Again this is due solely to the second ratio of $\Gamma$ functions in \tref{pkcomplete} which is present \emph{only} if the factors of $(1-\Pi)$ are included in \tref{neqngen}.

Only in the $E \ra \infty$ limit is the  transition between condensate and non-condensate regimes a phase transition and this occurs at $p_r=0$.

\begin{figure}[htb!]
\centering\includegraphics[width=6cm]{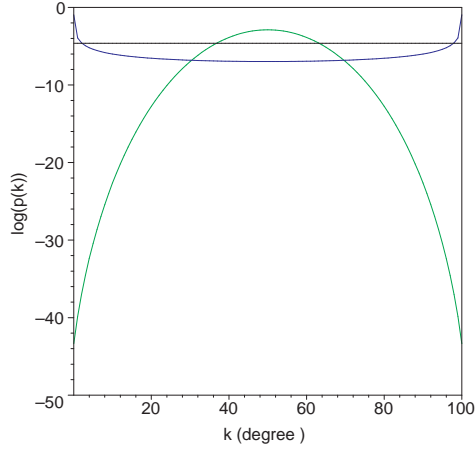}
\caption{Mean degree distribution for the Voter model for $E=100$ and various $p_r$.  The blue curve with the highest value at $k=0$ and $k=100$ is for $p_r=0.001 \ll E^{-1}$ shows a large consensus or condensate.  The horizontal black line is for $p_r=0.019 \approx p_\sharp$ and here there is a $Z_2$ transition between graphs with a minimum at $k=E/2$ to those with a maximum.  The green curve with lowest values at $k=0$ and $k=100$ is for $p_r=0.99$, well in the regime where there is no consensus.  Note that for $N=2$ there is a symmetry $p(k)=p(E-k)$.}
 \label{fvotertrans}
\end{figure}

A special case of interest is when we look at just two artifacts, $N=2$, so that $\kav$ is as large as possible.  With $p_r=0$ we obtain the basic Voter
model (see for example \cite{Liggett99,SR05}), which has been used as a simple model of language evolution \cite{SCES07}.  One question asked is the time for the model to come to a complete \emph{consensus}, i.e.\ all `voters' have made the same choice, a condensate in our language.  Our results show that the time scale is set by $\tau_2 := 1/\ln(\lambda_2)$.  A little randomness, $0 < p_r \lesssim E^{-1}$ leaves the consensus imperfect but still largely intact for very long periods of time.  This consensus will still take $O(E^2)$ rewirings to appear.  However for $p_\sharp = (1+(E/2))^{-1} \gg p_r \gtrsim E^{-1}$  while we still get most voters choosing the same option but the time scale for the equilibrium to be reached drops to $O(E)$ as $p_r$ is raised.  Finally there is a transition at $p_\sharp$, a $Z_2$ symmetry breaking transition,   to a region for large $p_r$ where there is no special consensus.  An example of some mean degree distributions in the Voter model with randomness added is shown in Fig. \ref{fvotertrans}.  These results are easily generalised to other large $\kav$ cases of the Copying Model e.g.\ we find that in general $p_\sharp = (E+1+\kav)^{-1}$.

\subsection{General Features of the Exact Solution}

The exact solution has several interesting properties.  The eigenfunction numbered zero is the only one which is time independent, $\lambda_0=1$, so this eigenfunction corresponds to the unique equilibrium solution. The eigenfunction numbered one \textit{never contributes}.  The first moment ($\lambda_1 \neq 1$) is constant as it is related to the ration $\kav = E/N$ yet it depends only on the eigenfunctions zero and one.  Since the later gives a time dependence its contribution must be zero, $c_1=0$.  Thus the slowest time dependence comes from $m=2$ eigenfunction,
setting the equilibration time scale to be
\beq
 \tau_2 = -1/ \ln(\lambda_2) \approx \frac{2p_r}{E}+ \frac{2(1-p_r)}{E^2} \, .
 \label{tau2def}
\eeq

The best way to study the time dependence is not to look at the moments but to study \emph{Homogeneity Measures} $F_n$
\bea
 F_n(t) &:=&   \frac{\Gamma(E+1-n)}{\Gamma(E+1)}
 \left. \frac{d^nG(z,t)}{dz^n}\right|_{z=1}
 \\
 &=&
  \sum_{k=0}^E \frac{k}{E}\frac{(k-1)}{(E-1)} \ldots
 \frac{(k-n+1)}{(E-n+1)} n(k,t)
 \; .
 \label{Fndef}
\eea
Trivially for $n>E$ $F_n=0$ but $\taverage{k^n} \neq 0$ highlighting one simplification over the moments.  It is also clear from the $(z-1)^m$ prefactor of the $m$-th eigenfunction \tref{Gmresult} that the $n$-th homogeneity measure
gets contributions only from eigenfunctions numbered $n$ and lower. The moments have a similar property as can be seen from the relationship between the $m$-th moments and the $n$-th Homogeneity Measures:
\bea
 F_n &:=& N \sum_{m=0}^n \Sone{n}{m} \taverage{k^m} \, , \;\;\;\;
 N \taverage{k^n} := \sum_{m=0}^n \Stwo{n}{m} F_m
\eea
where $\Sone{n}{m}$ and $\Stwo{n}{m}$ are Stirling numbers of the first and second kind respectively.  The generating function may now be
written as
\bea
 G(1+y,t) &=&  \sum_{n=0}^{E} y^n \binom{E}{n} F_n(t) =
\sum_{k=0}^E (1+y)^k n(k,t)
 \, ,
 \\
 F_n(t) &=&  \left. \frac{d^n G(z,t)}{d^nz}\right|_{z=1}
 \, ,
\eea
i.e.\ the $F_n$ are the $n$-th coefficients of the Taylor
expansion of $G$ around $(1+y)=z=1$.

\begin{figure}
\centering\includegraphics[width=7cm]{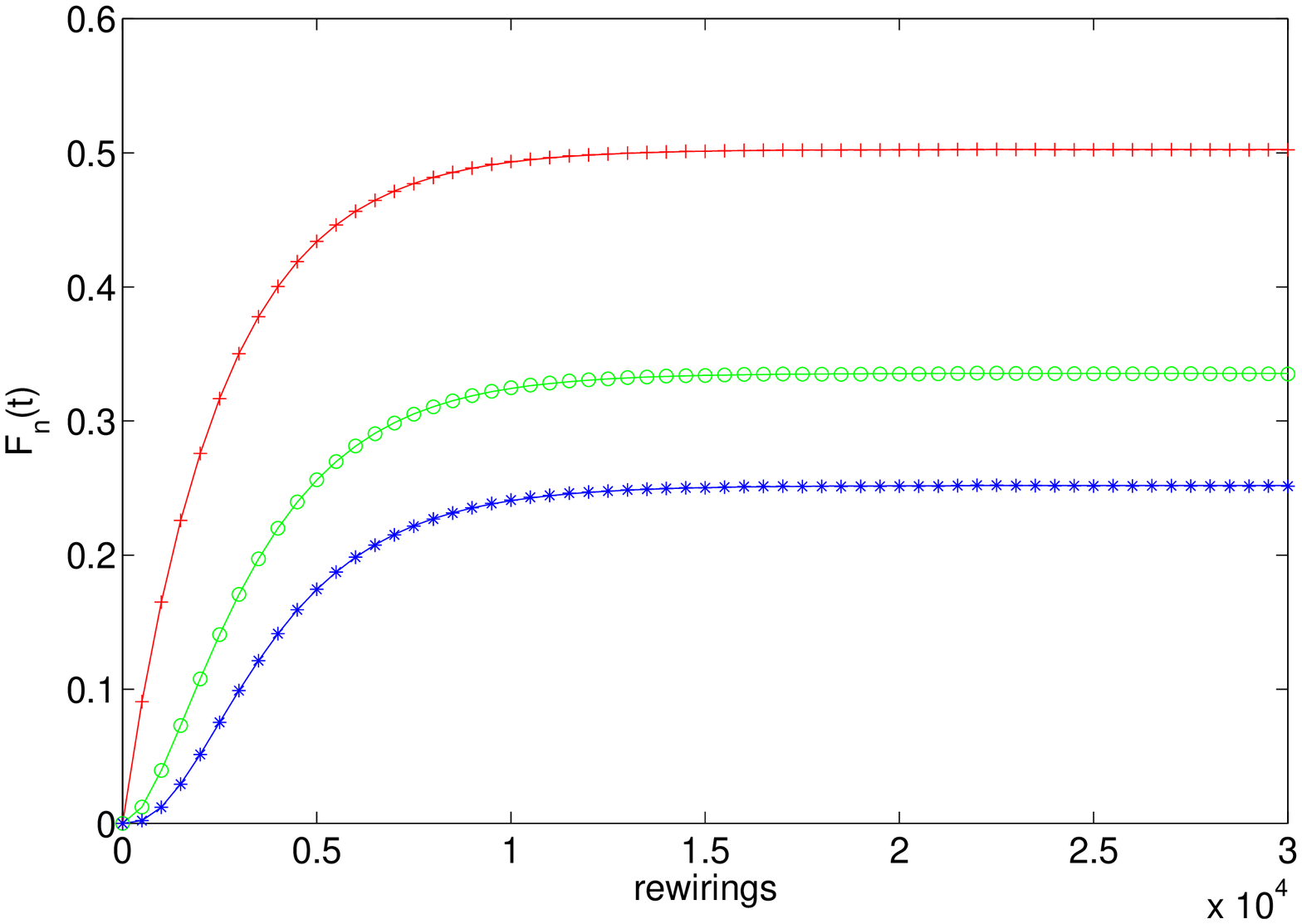}
\includegraphics[width=7cm]{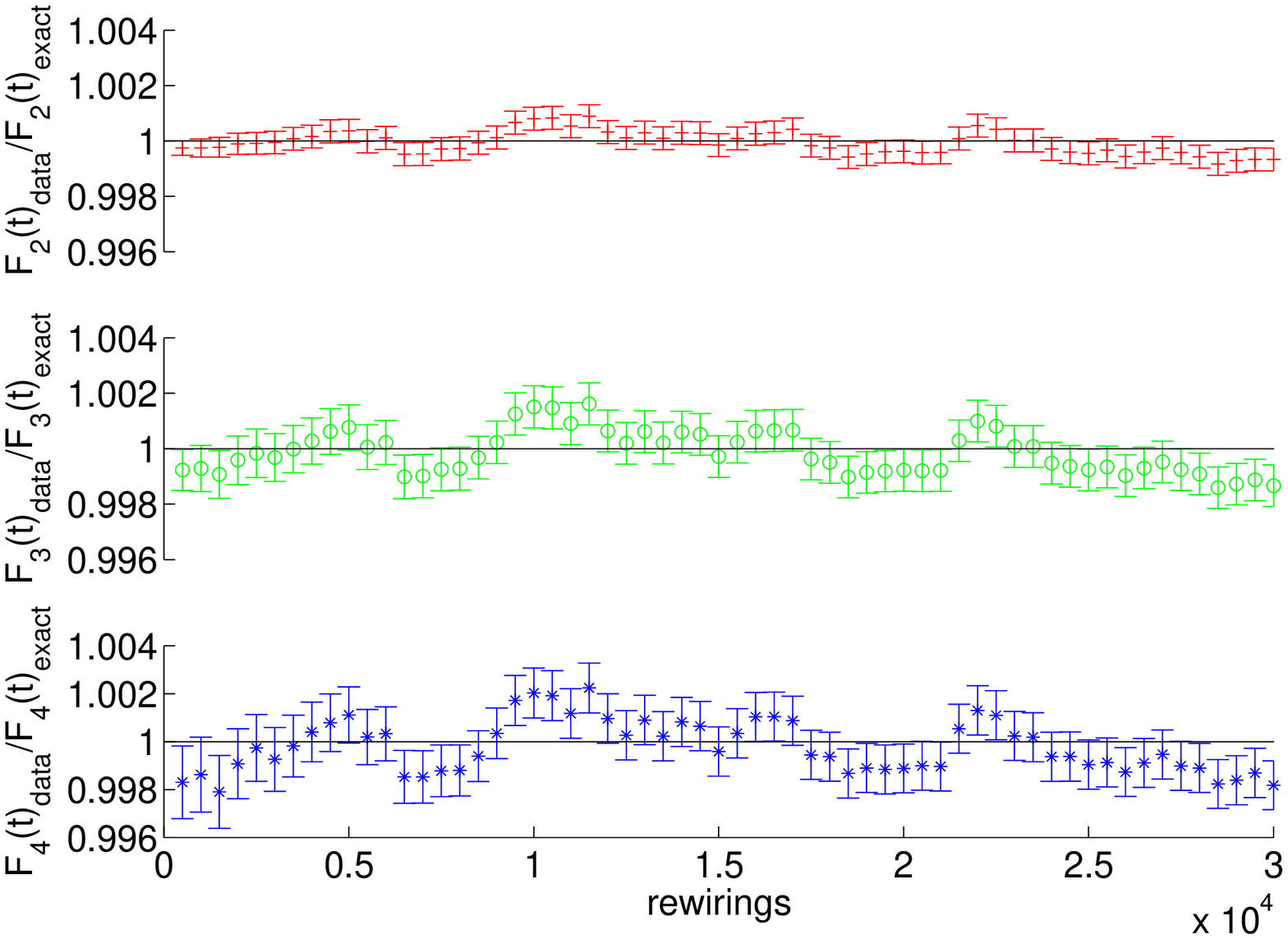}
\caption{Results for various $F_n(t)$ from numerical runs (points) with the lines given by the exact formula.  From top to bottom are: $F_2(t)$
(crosses), $F_3(t)$ (circles), $F_4(t)$ (stars). For $E=N=100$,
$p_r=0.01$ and data points are the average of $10^5$ runs of a
simulation. Taken from \cite{EP07}} \label{fnVarious}
\end{figure}

Unlike the moments, the homogeneity measures have a simple physical interpretation as they are the probability that any $n$ different individuals will have chosen the the same artifact.
Thus if for all $n\leq E $ we have $F_n=0$ then  no artifact has been chosen
more than once while if all $F_n=1$ then all individuals attached
to same artifact - a condensate.
For instance the simplest measure of homogeneity of the system is $F_2$, the probability that two different individuals have chosen the same artifact.  This is given by
\bea
 F_2(t) &=&
 F_2(\infty) + (\lambda_2)^t\left( F_2(0) - F_2(\infty) \right) \, ,
 \label{eqF2tres}
 \\
 F_2(\infty)
 &=&
 \frac{1+p_r(\langle k\rangle-1)}{1+p_r (E-1)} \, ,
\eea
where the initial conditions set $F_2(0)$.  The accuracy of the full time dependence of our solutions can be seen in following these $F_n$ meaures as shown in Fig. \ref{fnVarious}.

\subsection{Following a phase transition in real time}

\begin{figure}[hbt!]
\centering
\includegraphics[width=10cm]{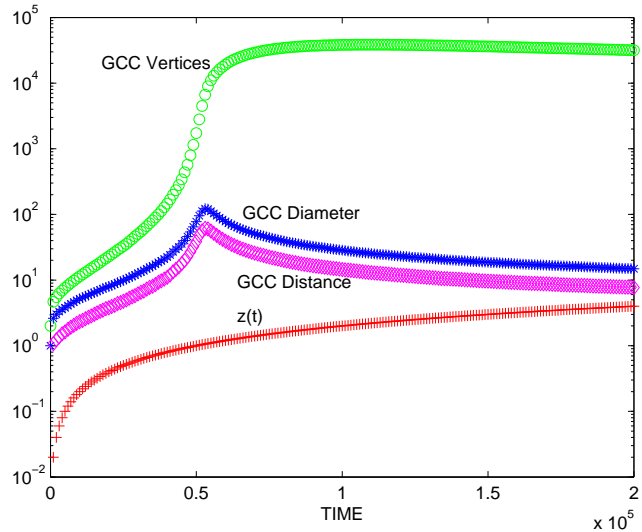}
\caption{Behaviour of a unipartite graph as a randomly end of a random edge is rewired at each time step, reattached to a vertex chosen with pure preferential attachment, $p_r=0$.  Here the graph has $N=E=10^5$ and starts from $F_2(0)=0$ i.e.\ all vertices are degree one and are connected to one other vertex.  Results are calculated for each instance and
then averaged over a total of 1000 runs.  Note finite size effects clearly visible as the transition is not perfectly sharp and it occurs at $z=1.06 \pm 0.01$.  Taken from \cite{EP07eccs07}.}
\label{fPTni1e5}
\end{figure}

We have already noted that our model gives the time evolution of the degree distribution of a generalised random graph made up of the artifact vertices, as shown in Fig. \ref{fCopyModel2und}. This graph undergoes a phase transition (e.g. appearance of GCC -
Giant Connected Component) at \cite{NSW01,DMS03a,FFH05,MR95}
$z(t)=1$ where $z$ was defined in \tref{zdef}.  This is simply related to $F_2(t)$ as
\beq
 z(t) := \frac{\ksqav   }{\kav} -1 = (E-1) F_2(t) \, .
  \label{zdef2}
\eeq
Thus we now have an analytic handle on the phase transition which occurs when rewiring a unipartite graph \cite{EP07eccs07}, shown in Fig. \tref{fPTni1e5}.  In principle we can calculate the number of vertices in the GCC, the diameter and average shortest path length in the GCC from known formulae and these only require knowledge of $F_2(t)$.

\subsection{Adding a Network of Individuals}

\begin{figure}[hbt!]
\centering
\includegraphics[width=8cm]{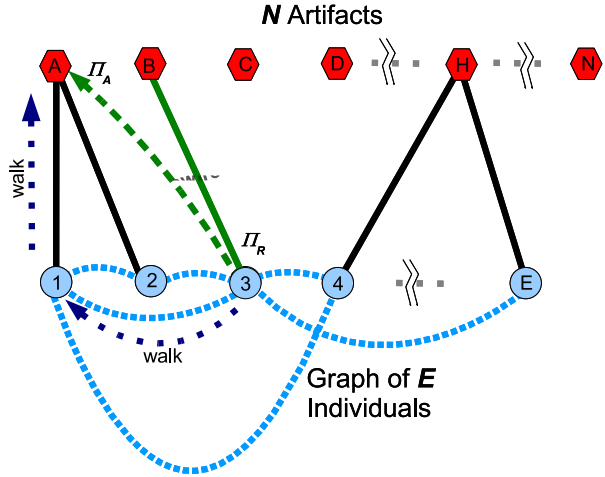}
\caption{The bipartite network of the Copying Model with an individual network added (shown below the individual vertices) and used to generate the copying mechanism.  The rewiring event is the same as that shown in Fig. \tref{fCopyModel3}, here assumed to be a copying event.  The target artifact is now found by first making a random walk on the individual network followed by a final step on the bipartite graph to reach the target artifact. In the case shown, the walk starts from from the randomly chosen individual 3 (its chosen artifact B is the source artifact) and we make a one step random walk on the individual graph to arrive at individual 1.  We finish by walking from individual one to its chosen artifact A, which becomes the target artifact.  Thus the event then consists of individual 3 copying the current choice of individual 1.}
\label{fcopymodel3ind}
\end{figure}

So far in this model we have inserted the copying process by hand, just demanding that the attachment probabilites $\Pi_A$ have a term proportional to $k^1$.  However, in practice we want to see this emerge as a natural process involving only local information.  Again the normalisation plays a key role since they contain global variables.  However it is simple to use the same random walk idea of section \ref{sslongtails} to generate the copying process in this model.  To do this we now an Individual graph, that is a network with edges between just the individual vertices, as shown in Fig. \ref{fcopymodel3ind}.

\begin{figure}[hbt!]
 \centering
 \includegraphics[width=8cm]{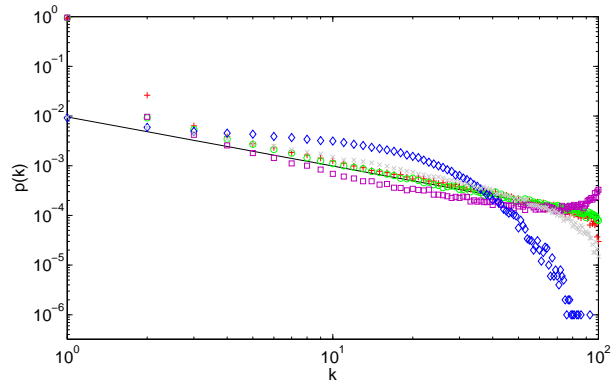}
\caption{Equilibrium artifact degree distribution $p(k)$ for
different Individual graphs of 100 vertices and average degree 4:
Erd\H{o}s-R\'{e}yni (red pluses), Exponential (green circles),
Barab\'asi-Albert (purple squares), periodic lattices of two (grey
crosses) and one (blue diamonds) dimension. The line is the analytic
result for a complete Individual graph while the other results are
taken over an ensemble of $10^4$ Individual graphs. $N=E=100$,
$p_r=1/E$. From \cite{EP07eccs07}.}
 \label{findnet}
\end{figure}

\begin{figure}[hbt!]
 \centering
\includegraphics[width=8cm]{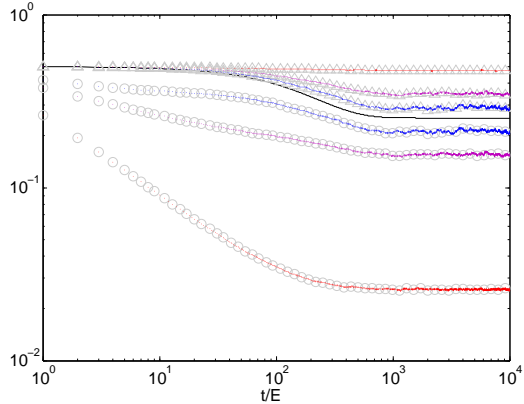}
\caption{Homogeneity measure $F_2$ for various cubic lattices against $t/E$.
The black solid line represents the analytic $ 1-F_2(t)$ for $N=2$,
$p_r=1/E$ and $E=729$. Numerical results for $1-F_2(t)$ (triangle
highlights) are plotted for 1-d (red), 2-d (purple) and 3-d (blue)
regular lattices. The average interface densities $\rhoexp$ are
plotted as circles. Averaged over $1000$ runs. From \cite{EP07eccs07}.}
 \label{flatcomp}
\end{figure}

The results suggest that in most cases the individual network has little effect on the equilibrium degree distribution \cite{EP07eccs07} as shown in Fig. \ref{findnet}\footnote{The lattices used in Fig.s \ref{findnet} and \ref{flatcomp} are periodic and cubic ($\mathbb{Z}^d$) with the nearest
neighbours connected except for one-dimension case where next-to-nearest neighbours are also connected. The
Exponential and Barab\'asi-Albert graphs are connected
Individual graphs with degree distributions of $p_\mathrm{ind}(k) \propto
\exp\{-\zeta k\}$ and $p_\mathrm{ind}(k) \propto [k(k+1)(k+2)]^{-1}$
respectively.}

However when we look at the time dependence we see more sensitivity to the properties of the individual graph.  We can now define a local measure of homogeneity, the average interface
density, $\rhoexp$.  This is the probability that any two individual vertices
which are connected by the Individual graph have a different artifact.
As Fig.~\ref{flatcomp} shows, the local and
global homogeneity measures $\rhoexp$ and $(1-F_2)$ are close to the
analytic result\footnote{The
analytic result equivalent to a complete individual graph with tadpoles, i.e.\ with adjacency matrix $A_{ij}=1$.} for large dimension lattices with short network
distances.  As we take lattices of smaller dimension, $F_2$ gets
much larger than the analytic result, and $\rhoexp$ much smaller.  Similar effects can be seen as we change $p_r$ \cite{EP07eccs07}.

These results are of relevance to many sociophysics models.
Consider a variation of the Minority game \cite{ATBK04} in which individual follow either their own strategy or that of a neighbour.  Then the number of `actors' (followers) using one particular strategy (that belonging to a leader) can be understood in terms of the Copying Model as this is the mean degree distribution with the strategies playing the roles of artifacts.  Actors are copying their strategy or, given the inherent instability of the game, they flounder about in a way that is statistically indistinguishable from a random (innovation) process.  It should be no surprise that this distribution is found to be \cite{ATBK04} a power law with slope one and some cutoff, in agreement with \tref{neqdist}. Again this model emphasises the way in which copying is a natural process in which preferential attachment and thus power law degree distributions emerge naturally, much as was noted for growing networks in \cite{SK04,ES05}.

\subsection{Different Update Methods}

\begin{figure}[hbt!]
\centering\includegraphics[width=8cm]{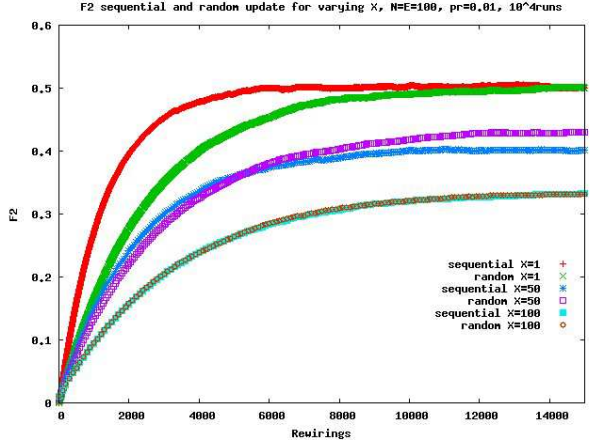}
\caption{Plots of $F_2$ for $E=N=100$ and $p_r=0.01$ for different update methods.  From top to bottom after 2000 rewirings we have sequential $X=1$ (red), random $X=1$ (green), sequential $X=50$ (blue), random $X=50$ (purple), and finally $X=100$.  Started from $n(k)=E\delta_{k 1}$ and averaged over $10^4$ runs. From \cite{EPY07}.}
\label{fupdateF2}
\end{figure}

One can also make changes to the way we update the system.  Suppose we first select $X$ different individuals at each step, either randomly or in numerical sequence (first $\{1,2,3,\ldots, X\}$ , then $\{ (X+1)  \mod E , (X+2)  \mod E, \ldots, (2X) \mod E\}$ etc.).  These individuals make their new artifact choices at the same time but still no updates occur. Finally the system is updated simultaneously. The simple Copying Model of \cite{Evans07,EP07eccs06,EP07,EP07eccs07} and discussed so far is the case of $X=1$ with random selection.  The models discussed in the context of cultural transmission \cite{Neiman95,BS03,HB03,HBH04,BHS04,BS05,BLHH07} choose $X=100$ where random and sequential updates are equivalent.  What we find numerically is shown in Fig. \ref{fupdateF2}.
When we update just one choice at each time step, $X=1$, sequential updating reaches equilibrium faster than random update but the equilibrium values are the same, $F_{2} =1/2$ in Fig. \ref{fupdateF2}.  This is to be expected as after $E$ updates random updating has not updated all elements while sequential has.  More surprising perhaps is the fact that for $X=E/2$ the time evolution is similar but sequential/random updates produce different equilibrium results.  This is also a lower equilibrium than the original $X=1$ model of \tref{neqngen} produces.  Finally we see that if we imitate \cite{Neiman95,BS03,HB03,HBH04,BHS04,BS05,BLHH07} and update all choices simultaneously, $X=E$, then we get the lowest equilibrium result, $F_{2} =1/3$ in Fig. \ref{fupdateF2}.

In fact it is possible to obtain analytic results for $F_n$ in the $X=E$ case and we find that we still have the same form $F_2(t) = F_2(0) + (\lambda_2)^t(F_2(\infty) - F_2(0))$ as  we found for $X=1$ random updating in \tref{eqF2tres} but now we have
\bea
 F_2(t=0, X=100) &=& \frac{p_p^2+ (1-p_p^2)\kav}{p_p^2+ (1-p_p^2)E} \, ,
 \label{F2X100}
 \\
 \lambda_2(X=100) &=& p_p^2\left(1- \frac{1}{E} \right) \, .
\eea
For the parameter values in Fig. \ref{fupdateF2} these formulae give the $F_2(\infty)$ values quoted above.

\subsection{Different Communities of Individuals}

Finally we can also look at a situation where we split the population into several communities.  Individuals in each community will then share the same copying and innovation probabilities, but now we are free to set different probabilities for copying which depend on the community of the individual being copied and on the community of the indivdual whose current choice is being copied.  What we wish to monitor is the number of times the different communities have chosen an artifact. Each artifact has a degree, $k_\alpha$, indicating how many times individuals from community $\alpha$ have chosen that artifact.  We therefore have to look at the mean degree distribution $n(\{k_\alpha\}; t)$. So the first step is to choose from which community the individual to be updated will be chosen and this can be done with probability $q_\alpha$ for community $\alpha$.  Once the source community $\alpha$ has been chosen, we then choose an individual at random from the $E_\alpha$ individuals in that community and it the choice of this individual that we are going to change.  This designates the source artifact which is about to lose an edge.  Now we have to determine the new choice for our chosen individual, the target artifact.  We can do this at random with some probability $p_{r\alpha}$, so that communities can have different innovation rates.  Alteratively the individual may decide to copy from an individual in some community $\beta$ which it does with probability $p_{p\alpha\beta}$.  In this process probability of attaching the edge being rewired, the choice made by an individual in community $\alpha$, will be proportional to the degree $k_\beta$ of each artifact.  For instance two extremes of behaviour would be when communities ignore the choices of other communities $p_{p\alpha\beta}=\delta_{\alpha\beta}$ or at the other extreme where all communities copy the `aspirational' choice made by a community $\gamma$ of `leaders' so $p_{p\alpha\beta}=\delta_{\beta\gamma}$. Within the limitation $p_{r\alpha}+ \sum_\beta p_{p\alpha\beta}=1$ there is a wide range alternatives.  For instance this might be suitable to model the choice of baby names which it has been suggested depends on the financial income of different groups \cite{Freakonomics}.

\begin{figure}[hbt!]
{\centerline{\includegraphics[width=8cm]{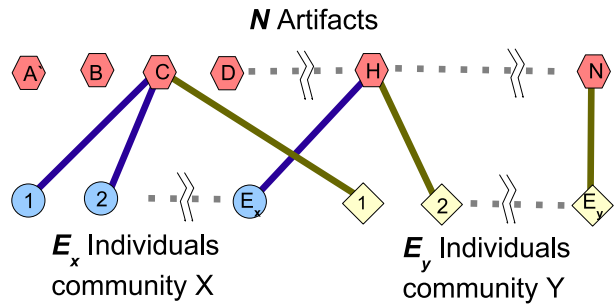}}}
\caption{The copying model with two communities of individuals, community X (Y) containing $E_x$ ($E_y$) individuals indicated as blue circles (yellow diamonds) with black edges (green edges) connecting them to the artifact that individual has chosen.}
 \label{fcopymodeltwocomm}
\end{figure}

With $C$ communities one finds a $(C+1)$-dimensional PDE for the generating function of $n(\{k_\alpha\}; t)$ but it does not admit an simple solution.  One may build solutions iteratively for $F(\alpha_1, \alpha_2, \ldots,\alpha_n;t)$, homogeneity measures which express the probability finding various types of edge attached to the same artifact.  Even then
the parameter space is now too large for a simple analysis.  For instance in the case of two communities `X' and `Y' ($\alpha=X,Y$) shown in Fig. \ref{fcopymodeltwocomm} there are eight parameters which may be chosen to be: $q_x, p_{pxx}, p_{pxy}, p_{pyx}, p_{pyy}, E_x, E_y$ and $N$.  The simplest homogeneity measures are $F_{XX}$, $F_{XY}$ and $F_{YY}$ where $F_{XX}$ is the probability that two different X individuals (individuals in community X) have chosen the same artifact, $F_{XY}$ is the probability that one randomly chosen X individual and one randomly chosen $Y$ individual have chosen the same artifact, and so forth.  These may be found analytically by finding the eigenvalues of a three-dimensional matrix but the details are lengthy and may be found in \cite{EP07eccs07}.

\section{Conclusions}

We have seen that a random walk is a very natural tool for analysis of generalised random graphs and for the analysis of real data sets. However it is much more than just an analysis tool.  Since a random walk can be performed using only local information it is also likely to be an important natural process in a wide variety of contexts.  If the use of a random walk is to find new and potentially better information from a network, then even if the actors in the system are only able to do short range walks, even if they can only look at their neighbours, then we are finding target vertices in roughly in proportion to their degree.  In the case of growing networks, this is the most natural way for preferential attachment to emerge and hence gives an explanation why so many power law degree distributions are found in data sets \cite{SK04,ES05}.

However we can take this a step further and imagine that having found a target vertex, we are likely to \emph{copy} some property of the  target.  In the growing networks model \cite{SK04,ES05} this was the creation of a new link to the target vertex by copying the target of an existing link, the end of the last edge followed in the random walk.  Thus preferential attachment processes can be seen as emerging from local searches done to exploit the information stored in the network, so that individuals/actors at a node may optimise their situation by learning from the knowledge represented by network.

The Copying Model is simplistic but because it captures such a basic and naturally emergent process --- copying --- on any sort of network, we should not be surprised to see it has such wide applicability.  For instance limiting ourselves to networks of constant size, its prediction of simple $1/k$ power laws with exponential cutoffs for certain parameter ranges means we can understand such laws in terms of this process when they are found elsewhere.  The simplest examples give us a rare example of an exactly solvable non-equilibrium process, known for any finite sized graph at for all times \cite{Evans07,EP07eccs06,EP07}.  However there are numerous extensions which may be needed for more realistic contexts where approximate analytical results may still be possible \cite{EP07eccs07}.

\section*{Acknowledgements}

I wish to thank Prof. E.Gelenbe, Prof S.Tucci and the staff of the Centro di Ricerca Matematica Ennio De Giorgi for organising such a stimulating workshop.  I would also like to thank A.Argent-Katwala, U.Harder, D.Hook, H.Morgan, A.D.K.Plato, J.Saram\"{a}ki, W.Swanell, D.Weir and T.You for their collaboration on various parts of the work described here.


\end{document}